
 \input phyzzx
 \def\abar{\overline{A}}
 \def\bd{B^0_d}
 \def\bdb{\overline{B}^0_d}

 \PHYSREV
 \doublespace
 \date{September 1992}
 \unnumberedchapters
 \pubnum{5911}
 \pubtype{T/E}
 \titlepage
 \title{Large Penguin effects in the CP Asymmetry of $B^0_d\rightarrow
 \pi^+\pi^-$
 \doeack}
 \author{Michael Gronau\foot{On leave of absence from the Physics
 Department, Technion, Haifa, Israel.}}
 \SLAC
 \abstract
Penguin effects in the CP asymmetries of $B^0_d\rightarrow \pi^+\pi^-$
, $\bd\rightarrow\rho^{\pm}\pi^{\mp}$ and $\bd\rightarrow a^{\pm}_1
\pi^{\mp}$ are studied as function of the CKM unitarity triangle
$\alpha$. Despite a fairly small penguin amplitude, it leads to quite
sizable uncertainties in the determination of $\sin(2\alpha)$ from all
but very large asymmetries. This effect is maximal for vanishing final
state interaction phases, for which it can cause, for instance, an
asymmetry of 40\%\ if $\alpha=\pi/2$.

 \submit{Physics Letters B}
 \endpage


There are two good reasons for which $B$ mesons provide a unique
opportunity for testing the Cabibbo-Kobayashi-Maskawa (CKM) mechanism
of CP violation%
\REF\nir{For a recent review see, Y. Nir and H.R. Quinn, SLAC-PUB-5737,
1992, to be published in Ann. Rev. Nucl. Part. Phys. We will use the
notations of this paper for the angles of the CKM unitarity triangle.}
\refend.
The CP asymmetries in certain decays, most notably decays to CP-
eigenstates, are expected to be both large and theoretically
clean.
For instance, in $B^0_d\rightarrow \psi K_S$ the time-dependent
asymmetry is predicted to oscillate with an amplitude given directly
by $\sin (2\beta)$%
\REF\sanda{A.B. Carter and A.I. Sanda, Phys. Rev. Lett. {\bf 45} (1980)
952; Phys. Rev. {\bf D23} (1981) 1587; I.I. Bigi and A.I. Sanda,
Nucl. Phys.{\bf B193} (1981) 85; {\bf B281} (1987) 41.}\refend ,
where $\beta$ is one of the angles of the CKM unitarity triangle.
This relation between a measured asymmetry and a pure CKM phase
parameter follows from having essentially a single
weak phase which contributes to the decay. In the case of $B^0_d
\rightarrow \psi K_S$, where it is known that
$\sin(2\beta)\ge 0.08$ \refmark\nir\ ,
this relation is expected to hold within a $1\%$ accuracy%
\REF\gronau{M. Gronau, Phys. Rev. Lett. {\bf 63} (1989) 1451.}\refend .

Another decay mode which seems to be very promising is $B^0_d\rightarrow
\pi^+\pi^-$,~in which the asymmetry is related to the angle $\alpha$.
In this case the theoretical situation is somewhat less clean due
to the contribution of ``penguin" amplitudes\refmark\gronau%
\REF\lon{D. London and R.D. Peccei, Phys. Lett. {\bf B233} (1989) 267.}
\refend\REF\grin{B. Grinstein, Phys. Lett. {\bf B229} (1989) 280.}
\refend , which may interfere with the dominant ``tree" amplitude
through their different weak phases. Here
the ratio of penguin-to-tree amplitudes is roughly estimated to be at
the level of $(10-20)\%$. This estimate
may lead one to conclude that an uncertainty at this level applies
also to the relation between the measured asymmetry and $\sin(2\alpha)$.
This by itself would not have spoiled the testing power of the
asymmetry measurement. The purpose of this note is to critically
elaborate on this question. We will show that, even with a relatively
small penguin contribution and with small final-state interaction
phases, penguin effects on the asymmetry may, in fact, be quite large
for $\vert\sin(2\alpha)\vert\leq 0.7$. Therefore,
unless a very large asymmetry is measured, this would prohibit
obtaining a useful value of $\alpha$ from the asymmetry measurement.

The formalism of studying CP asymmetries in neutral $B$ decays to CP
eigenstates in the presence of two interfering decay amplitudes was
given in \refmark\gronau\ . For completeness, we write down the basic
equations and study them for the case of $B^0_d\rightarrow \pi^+\pi^-$.
We denote the amplitude of $B^0_d\rightarrow \pi^+\pi^-$ by $A$ and that
of $\overline{B}^0_d\rightarrow \pi^+\pi^-$ by $\overline{A}$. Each of
these amplitudes obtains contributions from ``tree" and from ``penguin"
amplitudes:
$$
A=A_T e^{i\delta_T}e^{i\phi_T}+A_P e^{i\delta_P}e^{i\phi_P}~,
$$
$$
{}~\abar=A_T e^{i\delta_T}e^{-i\phi_T}+A_P e^{i\delta_P}e^{-i\phi_P}~.
\eqno\eq
$$
$A_{T,P}$ are real, $\phi_{T,P}$ are CKM phases and
$\delta_{T,P}$ are strong interaction final state phases, all
corresponding to the ``tree" and ``penguin" amplitudes, respectively.
It should be mentioned at this point that $\delta_{T,P}$ stand for
soft final state interaction phases. We neglect a phase due to the
absorptive part of the physical $c\overline{c}$ quark pair in the
penguin diagram%
\REF\soni{M. Bander, D. Silverman and A. Soni, Phys. Rev. Lett.
{\bf 43} (1979) 242.}\refend . This phase is very small at the
inclusive $b\rightarrow u\overline{u}d$ level%
\REF\chargedcp{G. Eilam, M. Gronau and J.L. Rosner, Phys. Rev. {\bf
D39} (1989) 819. In this paper this phase was estimated at ${\cal O}
(\alpha_s)$ to be about $3^0$. Even smaller phases were obtained at
${\cal O}(\alpha^2_s)$ by J-M Gerard and W-S Hou, Phys.
Rev. {\bf D43} (1991) 2909; H. Simma, G. Eilam and D. Wyler, Nucl.
Phys. {\bf B352} (1991) 367; See also L. Wolfenstein, Phys. Rev.
{\bf D43} (1991) 151.}\refend\ , and is not expected to be considerably
larger for exclusive modes such as $\pi^+\pi^-$, where the absorptive
part picks up contributions from a limited $q^2$ range%
\REF\simma{H. Simma and D. Wyler, Phys. Lett. {\bf B272} (1991) 395.}
\refend\ .

The time-dependent CP asymmetry for a neutral $B$ meson, created at $t=0$
as a pure $\bd$ and decaying at time $t$ to $\pi^+\pi^-$, when compared
to the corresponding decay rate of an initially $\bdb$, is
\refmark\gronau\ :
$$
Asym(t)=
{1-\vert{\abar\over A}\vert^2\over
1+\vert{\abar\over A}\vert^2}
\cos(\Delta mt)
-{2{\rm Im}({\abar\over A}e^{-2i\beta})\over 1+\vert{\abar\over A}
\vert^2}
\sin(\Delta mt)~.\eqno\eq
$$
$\Delta m$ is the mass-difference of the two neutral $B$ mesons.
The phase $2\beta$ appears in the $\bd-\bdb$ mixing amplitude
in the standard CKM phase convention.
The two terms in Eq.(2) describe two different kinds of CP violating
phenomena. The first $\cos(\Delta mt)$ term, which describes CP
violation in the direct decay of $\bd$, vanishes when only a single
CKM phase contributes to the decay process. The coefficient of the
well-known $\sin(\Delta mt)$ term, which appears when the mixed
$\bd$ and $\bdb$ decay to a comomn final state, is given by
$\sin(2\alpha)$  when only one  CKM phase contributes.

For $A_P/A_T\ll 1$ one finds the following expressions\refmark\gronau\
for the two coefficients in Eq.(2):
$$
{1-\vert{\abar\over A}\vert^2\over 1+\vert{\abar\over A}\vert^2}
\approx -2{A_P\over A_T}\sin(\phi_T-\phi_P)\sin(\delta_T-\delta_P)~,
$$
$$
{2{\rm Im}({\abar\over A}e^{-2i\beta})\over 1+\vert{\abar\over A}
\vert^2}\approx\sin(2\alpha)+2{A_P\over A_T}\sin(\phi_T-\phi_P)\cos
(2\alpha)\cos(\delta_T-\delta_P)~.\eqno\eq
$$
In order
to evaluate the penguin effects one must know the three quantities
$(\phi_T-\phi_P)$, $A_P/A_T,~(\delta_T-\delta_P)$. Only the first
quantity can be studied theoretically in a reliable manner. In the
standard CKM phase convention $\phi_T=phase(V^*_{ub}V_{ud})=\gamma$. The
penguin amplitude, on the other hand, obtains contributions from three
diagrams in which $u, c, t$ quarks run in a loop. Denoting the three
amplitudes, from which the CKM factors are omitted, by $P_u,~P_c,~P_t$,
one notes that within {\cal O}($m^2_c/m^2_b$) one has
$P_c\approx P_u$. It then follows from the unitarity of the CKM matrix
that
$\phi_P\approx phase(V^*_{tb}V_{td})=-\beta$, and therefore
$$
\phi_T-\phi_P\approx \gamma+\beta=\pi-\alpha~.
\eqno\eq
$$

The ratio $A_P/A_T$ cannot be calculated reliably at present. We will
attempt to evaluate it in two different manners. To be conservative,
we will try not to overestimate it. The ratio of loop-induced processes
$b\rightarrow dq\overline{q}~(q=u,d,s,c), dg, dgg$ to the tree process
$b\rightarrow u\overline{u}d$ was calculated perturbatively at the quark
and gluon level for the rates of inclusive charmless-strangeless
decays%
\REF\eilamgro{G. Eilam and M. Gronau, Phys. Rev. Lett. {\bf 61} (1988)
286; L. Lukaszuk, X.Y. Pham and C. Vu Xuan, Phys. Lett. {\bf B202} (1988)
388.}\refend . The penguin processes $b\rightarrow dq\overline{q}$
dominate the loop induced processes. The penguin-to-tree ratio of
rates was found to decrease as a function of $\vert V_{ub}/V_{cb}\vert$
within the range
$0.07<\vert V_{ub}/V_{cb}\vert<0.20$, from a largest possible value of
0.4 to a lowest value of 0.1, depending on $m_t$ and on CKM parameters.
The quark process $b\rightarrow du\overline{u}$ is likely to be the
dominant mechanism for $\bd\rightarrow \pi^+ \pi^-$. Its rate
is smaller by about 0.3 relative to all $b\rightarrow dq\overline{q}$
processes, which leads to a penguin-to-tree
ratio of amplitudes decreasing from 0.35 to 0.15 in the above range
of $\vert V_{ub}/V_{cb}\vert$. This may be a slight overestimate, since
in this calculation the penguin rate was maximized over the entire
acceptable CKM parameter space.

A somewhat different approach which leads to a similar estimate of
$A_P/A_T$ is based on calculating the low energy effective Hamiltonian
for $b\rightarrow d\overline{u}u$ in the leading log approximation
\refmark\grin\ . One finds that the penguin operators
$\overline{d}_{Li}\gamma^{\mu}b_{Lj}\overline{u}_{Lj}\gamma_{\mu}u_{Li}$,
{}~$\overline{d}_{Li}\gamma^{\mu}b_{Lj}\overline{u}_{Rj}\gamma_{\mu}
u_{Ri}$, ($i,j$ are color indices and $L,R$ are left and right
projections)
appear with coefficients 0.026, 0.033, respectively, and are
multiplied by the CKM factor $V_{tb}V^*_{td}$. On the other hand, the
relevant tree operator has a slightly enhanced coefficient ($=1.11$) and
is multiplied by $V_{ub}V^*_{ud}$.
Adding up the penguin coefficients and allowing the ratio $\vert
V_{tb}V^*_{td}/V_{ub}V^*_{ud}\vert$ to vary between the values of 1 and
5 \refmark\nir , one finds at the quark level $A_P/A_T\sim 0.05-0.27$.
Note that, since the operator coefficients are all
positive, $A_P/A_T$ (which does not include final state phases)
is positive too.

A very rough and oversimplified approximation, which represents
the above two results, somewhat on the low side, can be obtained
by simply using the CKM factors and the QCD factor related to the
single t-quark penguin diagram:
$$
{A_P\over A_T}\sim {\vert V^*_{tb}V_{td}\vert\over\vert V^*_{ub}V_{ud}
\vert} {\alpha_s\over 12\pi}\ln({m^2_t\over m^2_b})\sim 0.04-0.20~.
\eqno\eq
$$
The ratio of the CKM factors is the ratio of the lengths of two sides
of the CKM unitarity triangle which form the angle $\alpha$. As noted
above, this ratio lies between the values of 1 and 5 \refmark\nir\ .
Recent preliminary data%
\REF\CLEO{P. Drell, Talk at the XXVI International Conference on High
Energy Physics, Dallas, Texas, August 1992.}\refend\ , which seem to
indicate that $\vert V_{ub}/V_{cb}\vert$ is only about 0.06 or even
smaller, favor a large ratio.
For the QCD factor $(\alpha_s/12\pi)\ln(m^2_t/m^2_b)$ we took
the value 0.04, using $\alpha_s(m^2_b)\approx 0.2$. This value
would be larger if $\alpha_s$ were to be taken at $(m_b/2)^2$.

A large uncertainty is involved in calculating the tree and the penguin
operator matrix elements between the $\bd$ and the $\pi^+\pi^-$ states%
\REF\bsw{For a few models of two body $B$ decay matrix elements, see M.
Bauer, B. Stech and M. Wirbel, Z. Phys. {\bf C34} (1987) 103; M.
Tanimoto, Phys. Lett. {\bf B218} (1989) 481; A.
Szczepaniak, E.M. Henley and S.J. Brodsky, Phys. Lett. {\bf 243} (1990)
287.}\refend\ . Certain hadronic models seem to indicate that penguin
operator matrix elements may be enhanced due to their special chiral
structure%
\REF\shifman{M.A. Shifman, A.I. Vainshtein and V.I. Zakharov, Nucl. Phys.
{\bf B120} (1977) 315; M.B. Gavela {\it et al.}, Phys. Lett. {\bf 154B}
(1985) 425; N.G. Deshpande and J. Trampetic, Phys. Rev. {\bf D41} (1990)
895.}\refend . Since none of the existing methods of calculating
hadronic matrix elements is very reliable
for our case, we will make the most simplified assumption that
the ratio of these matrix elements is one, and will thus use Eq.(5) as
a crude approximation. This assumption has not yet been tested
experimentally even in an indirect way, that is, by comparing
tree-dominated to penguin-dominated processes. We feel that, since
the simplified relation (5)
somewhat underestimates the ratio calculated at the quark and
gluon level, it allows a certain amount of penguin matrix element
suppression, and does not overestimate the ratio of matrix elements.
We note again that
this ratio is likely to be on the high side of (5) if
$\vert V_{ub}/V_{cb}\vert$ is near its present lower limit value
of 0.06.

The soft final state interaction phase difference, $\delta_T-\delta_P$,
is basically uncalculable. Denoting this phase-difference by $\delta$,
one finds from (2)-(4):
$$\eqalign{
Asym(t)&\approx -2{A_P\over A_T}\sin\delta\sin\alpha{\bf cos}
{\bf (\Delta mt)}\cr\crr
&\quad-[\sin(2\alpha)+2{A_P\over A_T}\cos\delta\cos(2\alpha)\sin\alpha]
{\bf sin}{\bf (\Delta mt)}~.\cr}\eqno\eq
$$
We note that the $\cos(\Delta mt)$ term and the $\sin(\Delta mt)$ term
have a different and complementary $\delta$-dependence. The first term,
which describes CP violation in the direct decay, behaves like
$\sin\delta$, while the correction to the mixing-induced asymmetry is
proportional to $\cos\delta$. Thus, as function of $\delta$, the smaller
the direct CP violation $\cos(\Delta mt)$ term, the larger becomes the
penguin correction to $\sin(2\alpha)$, and vice versa. In particular,
when $\delta=0$, the $\cos(\Delta mt)$ term vanishes, whereas the
correction to the $\sin(2\alpha)$ coefficient becomes maximal.

A heuristic argument for factorization of tree amplitudes
in certain two body $B$ decays%
\REF\bj{J.D. Bjorken, in Proceedings of the Eighteenth
SLAC Summer Institute on Particle Physics, July 1990, ed. J. Hawthorne
(SLAC-Report-378), p. 167; M.J. Dugan and B. Grinstein,
Phys. Lett. {\bf B255} (1991) 583.}\refend\ implies that $\delta_T$ is
negligible and that perhaps also $\delta_P$ is small.
{\it If $\delta$ were small, then the $\cos(\Delta mt)$ term may be too
small to be observed and the time-dependent asymmetry measurement
would not provide evidence for a penguin contribution. Still, in this
case the penguin amplitude effect on the
coefficient of the $\sin(\Delta mt)$ asymmetry becomes maximal and
may be large.
This is the danger of penguin amplitudes}.

The crucial point is that, whereas one would naively expect that the
penguin amplitude modifies $\sin(2\alpha)$ in a multiplicative manner,
the correction is in fact an additive one and involves a factor of
two from the interference with the tree amplitude. This means that with
e.g. $A_P/A_T=0.2$, the correction to $\sin(2\alpha)$ can be as
large as $\pm 0.4$ and is not merely a relative $20\%$ correction. To be
quantitative, let us assume that $\delta$ is negligibly small, and study
the consequences of
(6) on the determination of $\sin(2\alpha)$ from an asymmetry
measurement. Fig. 1 shows the coefficient of the $-\sin(\Delta mt)$
term as function of the actual value of $\sin(2\alpha)$ for $45^0\leq
\alpha\leq 135^0$. The range bounded by the two solid lines describes
this coefficient for $\delta=0$ and for $A_P/A_T$ in the range (5). The
straight dashed line gives the corresponding relation in the absence
of a penguin contribution. The maximal deviation from the straight line
is given by $2A_P/A_T$. We note that points with largest deviations
 from this line correspond to the largest value of $A_P/A_T$ and thus
to the smallest values of $\vert V_{ub}/V_{cb}\vert$. The danger of the
penguin amplitude is best demonstrated for $\alpha=\pi/2$, where its
effect on the asymmetry is maximal. For this case, an asymmetry as large
as 0.4 can possibly be measured, although $\sin(2\alpha)=0$.
Such a substantial CP asymmetry measurement would be an important
observation by itself, however it could not be related to the angle
$\alpha$. The deviation of the asymmetry from $\sin(2\alpha)$ decreases
gradually, as one moves away from $\alpha=\pi/2$. It is still at a level
of $30\%$ at $\alpha=65^0,~115^0$ ($\sin(2\alpha)=+0.77,~-0.77$),
where one expects large asymmetries. The effects become much smaller
outside the range $45^0\leq \alpha\leq 135^0$ plotted in Fig. 1.
Unfortunately, the asymmetry measurement, which is related to the value
of $\sin(2\alpha)$, cannot distinguish between angles which lie outside
and inside this range. Note that as far as the CP asymmetry measurement
is concerned, the corrections are potentially large for all but very
large asymmetries. The range $65^0<\alpha< 115^0$, where
the corrections are larger than $30\%$%
\REF\comment{In general, there is a certain phenomenological correlation
between $\alpha$ and the allowed values of the ratio of the CKM factors
in (5)\refmark\nir\ . We use (5), although it seems that this ratio must
be somewhat smaller than 5 for $65^0\leq\alpha\leq 115^0$ if $\vert
V_{ub}/V_{cb}\vert>0.06$. The ratio of CKM factors may become larger with
smaller values of $\vert V_{ub}/V_{cb}\vert$ \refmark\CLEO\ .}\refend\ ,
is presently allowed\refmark\nir\  . It is possible that
future theoretical and experimental progress in quantities such as $f_B$
(the $B$ decay constant), $\vert V_{ub}/V_{cb}\vert$ and $m_t$ will rule
out this range%
\REF\rosner{J.L. Rosner, in B Decays, ed. S. Stone (World Scientific,
Singapore, 1992) p. 312.}\refend\ .

If one takes the very conservative viewpoint that the sign of the
penguin amplitude relative to the tree amplitude is unknown, then the
uncertainty in determining $\alpha$ from the asymmetry becomes twice as
large. Reversing the sign of $A_P/A_T$ corresponds to flipping the
allowed range for the asymmetry coefficient to the other side of the
straight dashed line. It would therefore be useful to at least
theoretically determine
the sign corresponding to $\delta\leq\pi/2$.
Since the penguin correction to $\sin(2\alpha)$ is proportional to
$\cos\delta$, it becomes substantially smaller than in Fig. 1 only for
large values of the final state phase difference. In this case one
expects to observe also the $\cos(\Delta mt)$ term in the time-dependent
asymmetry. This depends, of course, on the value of $\alpha$ and on the
sensitivity of the experiment. With no observation of such a term, and
without a theory of final state interaction phases, one would have to
assume the worst of all cases, namely $\delta\sim 0$ (or even
$\delta\sim\pi$, if the sign of the penguin amplitude is undetermined).

Penguin contributions appear also in $\bd\rightarrow \rho^{\pm}
\pi^{\mp}$ and in $\bd\rightarrow a^{\pm}_1\pi^{\mp}$.
The effects on a determination of $\alpha$ from the asymmetry
of the related time-dependent rates is expected to be as large as in
$\bd\rightarrow\pi^+\pi^-$. We wish to briefly demonstate this effect in
the decays to $\rho\pi$.
The general formalism dealing with this kind of final
states (which, although being non-CP-eigenstates, are common to $\bd$ and
$\bdb$ decays), was described in%
\REF\timedep{M. Gronau, Phys. Lett. {\bf B233} (1989) 479.}\refend .
The essential difference with respect to $\bd\rightarrow \pi^+\pi^-$ is
that here the tree and penguin amplitudes for $\bd\rightarrow \rho^+
\pi^-$ are not the same as those for $\bd\rightarrow\rho^-\pi^+$.
In both cases the corresponding amplitudes carry the same CKM
phases as in $\bd\rightarrow\pi^+\pi^-$,
$\phi_T=\gamma$ and $\phi_P=-\beta$, respectively:
$$
A_f\equiv A(\bd\rightarrow\rho^+\pi^-)=
A_T e^{i\delta_T}e^{i\phi_T}+A_P e^{i\delta_P}e^{i\phi_P}~,
$$
$$
A_{\overline{f}}\equiv A(\bd\rightarrow\rho^-\pi^+)=
\abar_T e^{i\overline{\delta}_T}e^{i\phi_T}+\abar_P e^{i\overline{
\delta}_P}e^{i\phi_P}~.\eqno\eq
$$
The corresponding amplitudes for the charge-conjugated processes,
$\abar_{\overline{f}}\equiv A(\bdb\rightarrow\rho^-\pi^+),~
\abar_f\equiv A(\bdb\rightarrow\rho^+\pi^-)$, are
obtained simply by changing the sign of the weak phases.

One can measure four different time-dependent decay rates, for cases in
which initially $\bd(\bdb)$ decay to $\rho^{\pm}\pi^{\mp}$. If $A_P$
could be neglected then these four rates would be, in principle,
sufficient for a determination of $\alpha$%
\REF\kayser{R. Aleksan, I. Dunietz, B. Kayser ad F. Le Diberder, Nucl.
Phys. {\bf B361} (1991) 141.}\refend .
As in decays to CP eigenstates, this method involves a measurement of
the coefficient of the $\sin(\Delta mt)$ term in the time-dependent rate.
For an initially $\bd$ decaying to $\rho^+\pi^-$ this coefficient is
given by
$$
{\rm Im}({\abar_f\over A_f}e^{-2i\beta})={\abar_T\over A_T}\sin(2\alpha+
\overline{\delta}_T-\delta_T)~.\eqno\eq
$$
If all final state phases were negligible, this coefficient would
determine $\sin(2\alpha)$. (Otherwise, the phase difference $\overline{
\delta}_T-\delta_T$ can be determined separately from the four rates).
In the presence of the penguin
amplitudes, this coefficient becomes, to lowest order in $A_P/A_T$ and
$\abar_P/\abar_T$, and for negligible final state phases,
$$
{\rm Im}({\abar_f\over A_f}e^{-2i\beta})={\abar_T\over A_T}[\sin(2\alpha)
+{A_P\over A_T}\sin(3\alpha)-{\abar_P\over\abar_T}\sin\alpha]~.\eqno\eq
$$
The correction to $\sin(2\alpha)$ can be estimated in a way similar
to the correction in $\bd\rightarrow \pi^+\pi^-$. Within our
approximation, Eq.(5) applies to both $A_P/A_T$ and $\abar_P/\abar_T$.
In fact, if one takes $A_P/A_T=\abar_P/\abar_T$ (which should hold only
for CP-eigentates) the correction term
obtains exactly the form of the correction term of Eq.(6). In general,
when these ratios are in the range (5), the effect of the penguin
amplitude on determining $\sin(2\alpha)$ are expected to be as large as
in $\bd\rightarrow \pi^+\pi^-$.

In summary, we have shown that relatively small penguin amplitudes may
prohibit a useful determination of $\sin(2\alpha)$ from the
CP asymmetries of $\bd\rightarrow \pi^+\pi^-$, $\bd\rightarrow
\rho^{\pm} \pi^{\mp}$ and $\bd\rightarrow a^{\pm}_1\pi^{\mp}$.
Asymmetries as large as 0.4 may be measured even when $\sin(2\alpha)=0$.
The uncertainty becomes small only for very large asymmeries. It would
decrease if $\vert V_{ub}/V_{cb}\vert$ were found to be on the high side
of the presently allowed range. The penguin complication
may be avoided to a large degree if future studies of the CKM matrix
exclude the range $65^0\leq\alpha\leq 115^0$ in which the corrections
are large. Our analysis was based primarily on the estimate (5) and on
the observation that the correction to $\sin(2\alpha)$ in (6) is additive
rather than multiplicative, and becomes maximal
when $\delta\rightarrow 0$. We assumed that the hadronic matrix element
of the penguin operator is neither dynamically suppressed nor enhanced
relative to the tree amplitude. It goes without saying that this issue
deserves serious studies, both theoretical and experimental. The mere
determination of the sign of the penguin amplitude would be useful.

One way to overcome this potential difficulty is to
measure in addition to the asymmetry in $\bd\rightarrow\pi^+\pi^-$ also
the rates of $\bd\rightarrow\pi^0\pi^0,~B^+\rightarrow\pi^+\pi^0$.
This isospin-based method%
\REF\grolon{M. Gronau and D. London, Phys. Rev. Lett. {\bf 65} (1990)
3381.}\refend
can provide a way to eliminate the penguin contribution altogether
and to experimentally determine its magnitude,
provided that the integrated rate into two neutral pions is measurable.
A similar isospin analysis for the $\rho\pi$ modes is unlikely to work
in practice due to the too many amplitudes involved
and to certain ambiguities which appear in the analysis%
\REF\quinn{H.J. Lipkin, Y. Nir, H.R. Quinn and A. Snyder, Phys. Rev.
{\bf D44} (1991) 1454; M. Gronau, Phys. Lett. {\bf B265} (1991) 389.}
\refend .

I wish to thank J. Bjorken and H. Quinn for useful discussions.
\endpage

\refout
\endpage
\centerline{\bf FIGURE CAPTION}
\item{\rm FIG.1.} Coefficient of $-\sin(\Delta mt)$ in the asymmetry
of $\bd\rightarrow\pi^+\pi^-$
as function of $\sin(2\alpha)$ for $45^0\leq\alpha\leq 135^0$. We take
$\delta=0$. Area between solid lines corresponds to $A_P/A_T=0.04-0.20$;
dashed line corresponds to the absence of a penguin amplitude.

 \bye